# Nanoparticles Binding to Lipid Membranes: from Vesicle-Based Gels to Vesicle Inversion and Destruction

Derek A. Wood[a], Sarah Zuraw-Weston[a], Ian K. Torres[a], YiWei Lee[b], Li-Sheng Wang[b], Ziwen Jiang[b], Guillermo R. Lázaro[c], ShiYu Wang[d], Avital A. Rodal[d], Michael F. Hagan[c], Vincent M. Rotello[b], and Anthony D. Dinsmore[a]

[a]Department of Physics, University of Massachusetts Amherst; [b]Department of Chemistry, University of Massachusetts Amherst; [c]Department of Physics, Brandeis University; [d]Department of Biology, Brandeis University

May, 2018.

**ABSTRACT**

**Cells offer numerous inspiring examples where proteins and membranes combine to form complex structures that are key to intracellular compartmentalization, cargo transport, and specialization of cell morphology. Despite this wealth of examples, we still lack the design principles to control membrane morphology in synthetic systems. Here we show that even the relatively simple case of spherical nanoparticles binding to lipid-bilayer membrane vesicles results in a remarkably rich set of morphologies that can be controlled quantitatively *via* the particle binding energy. We find that when the binding energy is weak relative to a characteristic membrane-bending energy, the vesicles adhere to one another and form a soft solid, which could be used as a useful platform for controlled release. When the binding energy is larger, the vesicles undergo a remarkable destruction process consisting first of invaginated tubules, followed by vesicles turning inside-out, yielding a network of nanoparticle-membrane tubules. We propose that the crossover from one behavior to the other is triggered by the transition from partial to complete wrapping of nanoparticles. This model is confirmed by computer simulations and by quantitative estimates of the binding energy. These findings open the door to a new class of vesicle-based, closed-cell gels that are more than 99% water and can encapsulate and release on demand. Our results also show how to intentionally drive dramatic shape changes in vesicles as a step toward shape-responsive particles. Finally, they help us to unify the wide range of previously observed responses of vesicles and cells to added nanoparticles.**

Significance Statement:
Cells provide inspiring examples of controlling membrane deformation by means of interactions with viruses or proteins. Learning how to control this process in a synthetic system will open up new materials and also reveal mechanisms at play in live cells. In this work, we show how tuning the strength of binding between spherical nanoparticles and membranes controls the membrane deformation process. Responses range from membrane-membrane adhesion (leading to a new form of closed-cell gel) to a remarkable vesicle-inversion and destruction process. Combined with computer simulations, these results show how the adhesion energy controls the behavior of vesicles, providing a major step forward in understanding how to tune membrane morphology in cells or in synthetic systems that mimic cells.

Keywords:
Keywords: membrane morphology; lipid bilayer membrane; membrane physics; nanoparticle-membrane interactions; membrane nanotube

The lipid-bilayer membrane plays a central role in biological processes owing to its properties as a thin layer that is flexible, impermeable to most solutes, and fluid-like in its plane (1, 2). The membrane's ability to adopt complex shapes such as in the endoplasmic reticulum is important for protein synthesis and transport, and its ability to change shape plays a crucial role in forming protrusions for cell mobility (3) or engulfing objects in endocytosis (4, 5). These cell-based examples have inspired the application of synthetic lipid bilayers for encapsulation and delivery, but there is considerably greater (and still undeveloped) potential if we can learn how to trigger changes in membrane geometry and topology. We seek to understand how to mimic this functionality with the ultimate aim of constructing new responsive, bioinspired materials that can modulate morphology and function on demand.

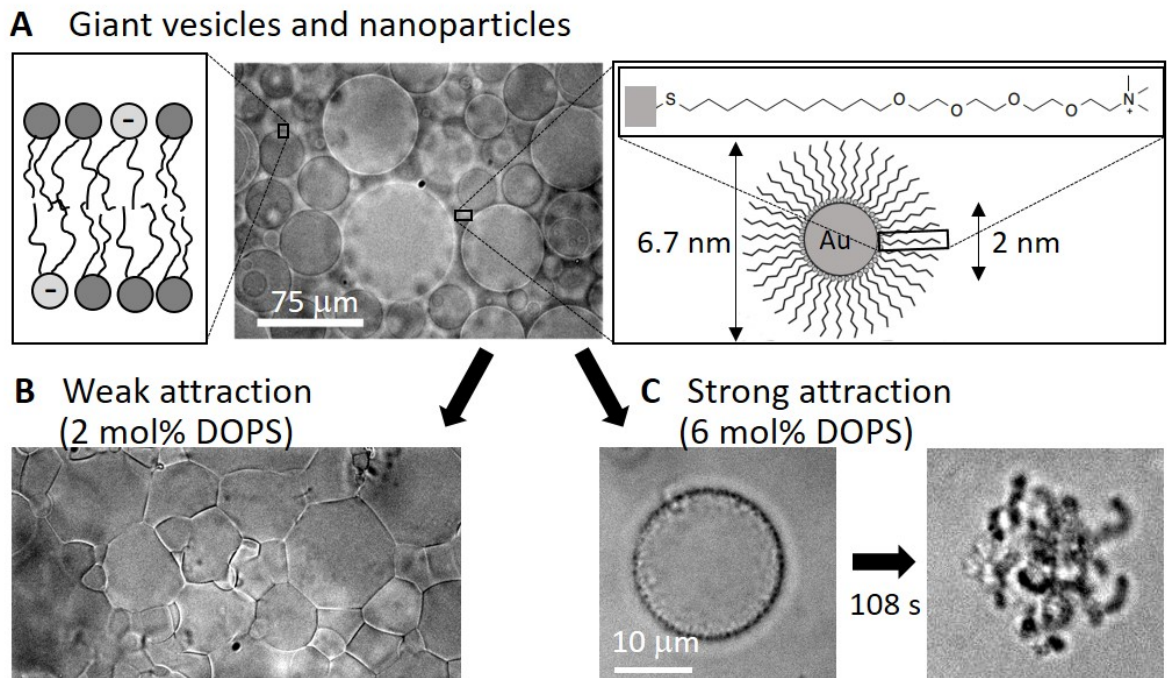

**Fig. 1.** (*A*) Schematic overview of giant unilamellar vesicles (GUVs) with controllable anionic charge density exposed to cationic gold nanoparticles (Au-TTMA). The microscope image shows GUVs composed of 96 mol% DOPC and 4 mol% anionic DOPS without nanoparticles. (*B*) Microscope image of GUVs + NPs that have adhered to one another forming a solid gel. (*C*) Microscope images of a single GUV undergoing rapid shape inversion and destruction upon Au-TTMA nanoparticle binding.

The deformation of a membrane upon binding of a particle or macromolecule is currently understood as a competition between the free-energy reduction from binding and the free-energy increase from bending the membrane, which is treated as a continuous elastic body (6-8). Defining the binding energy per unit area of contact as $w$, the radius of a particle or virus or macromolecule as $a$ and the membrane bending modulus (8) as $\kappa$, a crossover from mild deformation to full wrapping of the membrane around the bound object is expected when $wa^2/\kappa$ is on the order of 1 (6, 9). In the regime of individual particles, calculations and simulations (6, 9-12) and experiments support this picture (7, 13-15). When many particles or viruses are present, experiments show that cooperative interactions lead to in-plane attraction between two particles (14), clustering (16, 17), tubulation or pearling of the membrane (18-21), and internalization of particles within the vesicles (7, 13). Similarly, simulations and calculations found hexagonal or chain-like particle aggregates (22, 23), budding or tubulation of the membrane (11, 21, 24-29), or internalization (10, 30). Additional work has shown that bound nanoparticles build additional functionality into vesicles, including the ability to lyse and release cargo through UV light-induced heating of adsorbed particles (31). Despite the wide range of phenomenology, it is still not known how to predict or control these collective particle-membrane behaviors – information that is needed to rationally tune membrane shape for applications, or to understand how cells carry out this task.

Here we report the results of a well-defined lipid membrane and nanoparticle system that allows us to tune the interaction strength, $w$, between the two components. We used giant lipid bilayer vesicles (10-100 μm) and 6.7-nm-diameter cationic Au-TTMA nanoparticles (Fig. 1*A*) (32, 33). The binding energy of the nanoparticles and vesicles was controlled *via* the ratio of two lipid species in the vesicle membranes: zwitterionic DOPC and anionic DOPS. When $wa^2/\kappa$ was small, the nanoparticles caused the vesicles to adhere to one another and form a soft but solid gel (Fig. 1*B*). By contrast, when $wa^2/\kappa$ exceeded a threshold value (Fig. S2), the vesicles were turned inside-out and utterly destroyed (Fig. 1*C*). These two behaviors can be explained at the microscopic scale by a transition from a partial wrapping of nanoparticles to complete wrapping by the membrane when $wa^2/\kappa$ exceeded a threshold value. By contrast, when non-particulate cationic polymers bound to the vesicle, we always found adhesion and gel formation with no vesicle disruption, showing that the rigid shape of the particles is necessary for membrane disruption.

The ability to tune the morphology and shape-changing dynamics of vesicles provides a useful experimental model of cell lysis and opens the door to applications. These findings could be used to create cargo-carrying vesicles with the ability to rupture on trigger, or to engineer soft solid gels from semi-permeable materials that can encapsulate cargo. They may also provide a unified picture for the wide variety of phenomena that have been reported in cells and vesicles, which likely correspond to different regions of a phase space defined chiefly by *w*, *κ*, *a*, and particle concentration.

**Results**
**Overview of the phenomenology.** By adjusting the anionic DOPS content of the vesicles, we tuned the average surface charge of the vesicles and thereby the adhesion energy per area, *w*, between the cationic nanoparticles and the lipid bilayer. We took care to keep the osmotic strength of the exterior solution the same as that of the interior solution, so that osmotic shock did not play a role in these processes. (See Methods section and SI for details.) Figure 1*B*,*C* summarizes the two distinct behaviors that we observed: adhesion and vesicle-gel formation at low DOPS fraction, and vesicle inversion and destruction at high DOPS fraction. Remarkably, these two regimes of behavior were separated by a well-defined threshold DOPS fraction. This threshold value increased from approximately 4 mol% to 5 mol% as the salt concentration was raised from no added salt to 20 mM NaCl. Below, we attribute this change to screening of the electrostatic attraction, which means that more DOPS is needed to achieve the same binding energy.

When the DOPS content in the membrane was < 4 mol%, the nanoparticles bound to the vesicles' surfaces without any discernible deformation. Evidence of particle binding is provided by dark-field optical microscopy, which showed enhanced scattering of light by the bound nanoparticles (Fig. S3). When the concentration of vesicles was high enough, the surfaces of nearby vesicles adhered to one another owing to the nanoparticles' forming an adhesive bridge between them. This adhesion process led to a solid network of vesicles. In overall appearance, the approximately polyhedral vesicles resembled bubbles in a dry soap foam, except that here the entire system was entirely aqueous. This closed-cell structure allows the gel to encapsulate a large volume of liquid within a series of robust interior partitions, forming a useful delivery vehicle for drugs, dyes, or reagents.

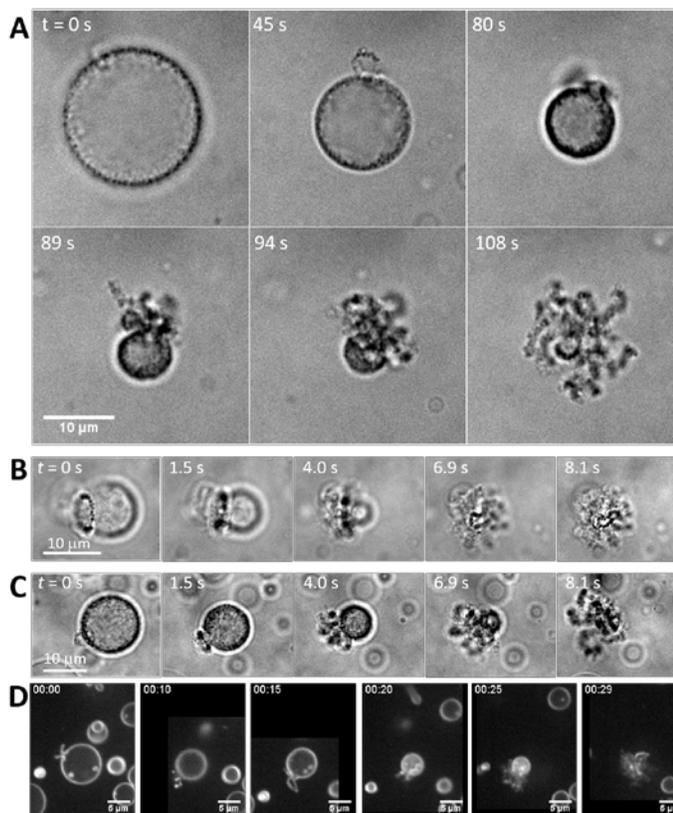

**Fig. 2**. A series of frames showing the time evolution of vesicles leading up to complete nanoparticle-induced disruption. (*A-C*) Bright-field images of 6 mol% DOPS; (*D*) Confocal fluorescence images of 5 mol% DOPS with <1 mol% Rh-DOPE (time in seconds).

By contrast, when the DOPS content achieved a threshold value (approx. 4 mol%), binding of the nanoparticles caused the vesicles to be completely disrupted in a remarkable, multi-stage process (Fig. 2). Although each vesicle differed in detail, the stages were common across hundreds of vesicles in dozens of different samples. We give a brief synopsis here and provide details in the next section. First, the diameter of the vesicle steadily decreased over a typical duration of several seconds to minutes as the membrane became loaded with nanoparticles (Fig. 3). During this stage, the vesicle developed dark spots that diffused on the surface (Fig. 4). In vesicles that shrank faster than a rate of 300 $\mu m^2$/s, vesicles also developed a single, long-lived pore in the membrane (Figs. 2,3,5). Remarkably, these pores were stable and had diameters in the range of 1-10 $\mu m$, much larger than the size of an individual nanoparticle. Finally, these vesicles underwent a complete inversion, where the interior of the vesicle was forced outwards through the pore (Fig. 2). Alternatively, vesicles that had not formed a pore suddenly ruptured and inverted. In all cases, the final condition appeared to be a network of tubule-shaped lipid bilayers coated with nanoparticles (final frames of Fig. 2). The supplemental movie shows this process in dark-field microscopy. In multilamellar vesicles, the outer layers of the vesicle peeled off one by one as they were attacked by the nanoparticles, until only one inner layer remained (Fig. S4).

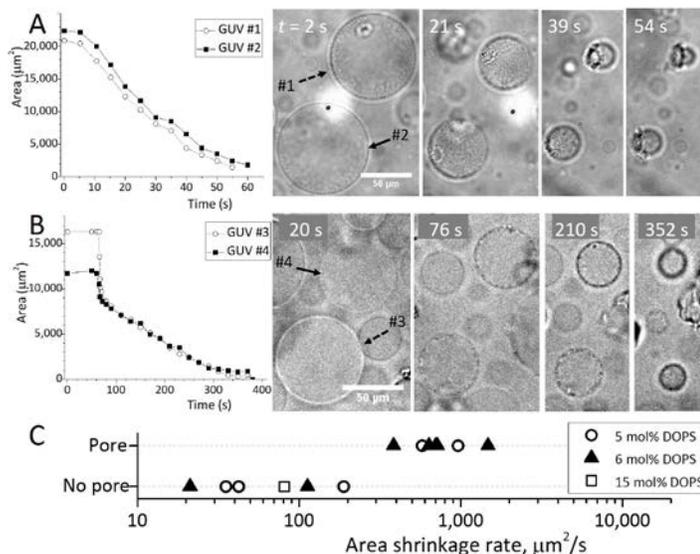

**Fig. 3.** Measured surface areas over time for vesicles attacked by nanoparticles. (*A*) 5 mol% DOPS; nanoparticles were added at *t* = -5 min. The average rate of area reduction was 500 $\mu m^2$/s. Both vesicles developed a surface pore (visible at *t* = 21 s), then gradually inverted through the pore as they shrank. (*B*) 5 mol% DOPS; nanoparticles were added farther away, at *t* = -50 min. and the local concentration of nanoparticles was lower than in (*A*). The average rate of area reduction was approx. 40 $\mu m^2$/s. The vesicles suddenly ruptured at *t* ≈ 400 s without having first formed a visible pore. (*C*) A plot of area shrinkage rates of 13 vesicles and various DOPS composition. All vesicles that shrank faster than 300 $\mu m^2$/s developed a pore.

In the following sections, we describe the phenomenology of the two separate regimes in greater detail. We then describe molecular dynamics simulations that show a similar crossover from weak binding to destruction. Then, in the Discussion section we describe the underlying mechanisms and compare to prior work.

**The stages of destruction (strong binding)**
Here we provide a more detailed description of the inversion and destruction process that occurs when the binding energy exceeds the threshold value. At the start of the disruption process, the diameter of the vesicle steadily decreased. As shown in Fig. 3, vesicles close to one another tended to shrink at similar rates. Fig. 3*A* shows vesicles that were close to the site of nanoparticle addition; for most of the shrinkage process, these two vesicles lost apparent surface area at an average rate of approx. 500 $\mu m^2$/s.

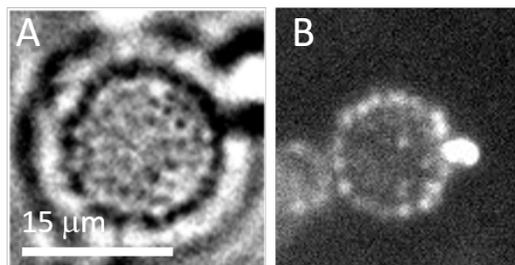

**Fig. 4.** Vesicles showing surface spots. (*A*) Bright-field image; 6 mol% DOPS. (*B*) Dark-field image highlighting nano-Au clusters; 4 mol% DOPS.

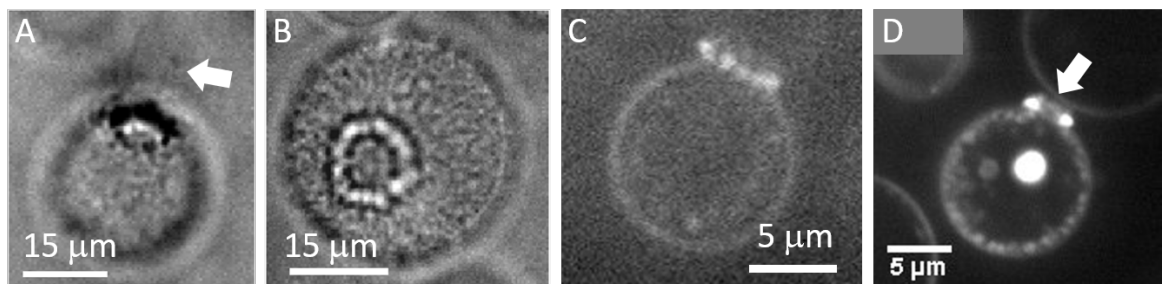

**Fig. 5.** Images of vesicles with a long-lasting pore. (*A-B*) Bright-field images, 6 mol% DOPS. In (*A*), fluid can be seen escaping the pore, as indicated by white arrow. (*C*) Dark-field image, 4 mol% DOPS. (*D*) Confocal image, 5 mol% DOPS + 1 mol% Rh-DOPE.

Surprisingly, the appearance of a large pore on the surface of each vesicle had no measurable impact on the shrinkage rate. Fig. 3*B* shows vesicles that were farther from the point of nanoparticle addition, so that the local nanoparticle concentration was reduced by diffusion over a period of 50 min prior to the start of shrinkage. The sharp initial decrease in radius observed for the vesicles in Fig. 3*B* was observed in many vesicles and is attributed to excess area in the initial configuration. Following this rapid decrease, the steady area-shrinkage rates were approx. 35 $\mu m^2$/s. This rate was approx. 14× lower than in Fig. 3*A*. In separate experiments, we added nanoparticles with a 14× reduced concentration and found that the average shrinkage rate decreased to 0.004 $\mu m^2$/s, and the rupture process required an hour or more to complete. These data show that the rate of vesicle shrinkage was strongly correlated with nanoparticle concentration. This point will be discussed below.

As the diameters of the vesicles shrank, dark spots appeared on the surfaces (Fig. 4). This effect was universal, with every vesicle imaged showing these dark spots in conjunction with surface shrinking. Because they were bright under dark-field imaging (Fig. 4*B*), we conclude that the spots were enriched in Au-TTMA nanoparticles, which implies an attractive interaction between particles mediated by the deformed membrane. These small dark spots were always similar in size to the microscope's resolution limit, so that their true size could not be measured accurately. However, their visibility indicates they were clusters of many nanoparticles and not individual nanoparticles. Furthermore, as more nanoparticles bound and the vesicle shrank, these dark spots visibly increased in number but did not increase in size. Throughout, they remained mobile on the vesicles' surfaces.

The formation of an open, micron-sized pore that persisted for at least several seconds is a very striking and unique feature of our results. Figures 5, S5, and S6 show that these pores are truly open. In Fig. 5*A*, escape of the encapsulated fluid (175 mOsm/L sucrose) can be seen because it has a different index of refraction than the exterior fluid (87.5 mOsm/L sucrose + 90 mOsm/L glucose), leading to a visible fingering effect. Furthermore, the confocal image Fig. 5*D* shows an open hole. A time-series of images of this vesicle shows that the bright lipid particle inside the vesicle was pushed out through that pore (see SI, Fig. S6). We found a characteristic 'pearl necklace' morphology at the outer rim of each pore, consisting of clearly discernible clusters that surrounded the rim of the pore. The dark-field image of Fig. 5*C* shows that these clusters were enriched in nanoparticles, visible by their strong scattering.

Approximately half of the imaged vesicles with ≥ 5 mol% DOPS formed a visible pore prior to the final inversion stage. A shown in Fig. 3*C*, only vesicles whose surface area decreased faster than a rate of approximately 300 $\mu m^2$/s formed a visible pore, regardless of the DOPS content of the vesicle (as long as it was above the threshold). The particle concentration determined the rate of vesicle shrinkage and, in turn, controlled pore formation.

The final stage of the disruption process was the complete inversion (turning inside-out) of the vesicle structure, resulting in a network of tube-like structures. Confocal microscope images show lipid-nanoparticle tubules inside the vesicle (Fig. S7), which emerged during the inversion (Fig. 2). From the optical images, we estimate that tubules had a typical diameter of approximately 1-2 $\mu m$. We found no evidence that the initial vesicle size or DOPS content (as long as it was above the threshold) affected the rate of shrinking of the vesicles or the sizes of the remaining tubule structures.

**Vesicle adhesion and gel formation (weak binding)** In the regime where the DOPS content of vesicles was less than the threshold value (approx. 4 mol%), the vesicles adhered to one another and formed a macroscopic network. Compared with the complex disruption process described in the previous section, this phenomenon is relatively straightforward: nanoparticles bound to the membrane and diffused laterally. When two vesicles came into contact, the nanoparticles bound to both of them and thus created an adhesive bridge. Between pairs of vesicles, the contact area grew over a typical time on the order a few minutes before reaching a steady state. A time-lapse series of images is provided in Fig. S8. Adhesion ultimately led to the formation of a macroscopic, gel-like aggregate of vesicles. Figure 6*A* shows a dark-field image of a steady-state gel, composed of pure DOPC vesicles (no DOPS). The strong light scattering at the adhesion sites between vesicles shows that Au-TTMA nanoparticles were enriched at these sites. This enhanced concentration can be explained by the roughly 2× greater binding energy owing to the two nearby membranes. No systematic variation of morphology was found in images of samples where the DOPS fraction varied between 0 and 3 mol%.

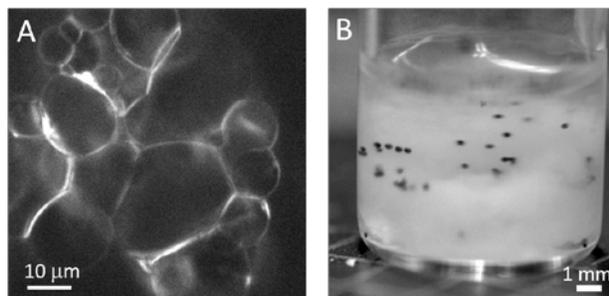

**Fig. 6.** (*A*) Dark-field image of a vesicle gel, showing nanoparticles in the adhesion regions (DOPC-only). (*B*) Photograph of a macroscopic gel composed of soy-lecithin PC vesicles with polycation (PDADMAC) added. The gel could support 300-μm-diam copper beads (dark spots), indicating a finite shear modulus and solid behavior.

To probe the mechanical properties of the vesicle-based gels, we developed an alternative system that can be made in large quantity using inexpensive, food-grade soy lecithin phosphocholine lipid (SLPC). Success in making large (50-mL) quantities shows the potential of this method for widespread application. To further expand the range of materials that can be used to form the gel, we added cationic polymer to induce vesicle aggregation without particle-induced destruction. We experimented with two polycations: poly-L-lysine (Sigma Aldrich, 150 kDa) and the more highly charged polydiallyldimethylammonium chloride (PDADMAC, 200 kDa). Each successfully caused aggregation of the vesicles into a gel. In all cases with polycations, we observed vesicle-vesicle adhesion and gel formation. Even with up to 15 mol% DOPS, we never observed the destruction process, which indicates that the rigid particle shape is necessary to trigger the destruction. Figure 6*B* shows a 0.5 mL-sample of SLPC-vesicle gel with PDADMAC. Copper beads of diameter 300 μm were added to the suspension and were clearly supported against gravity. The support of these beads indicates that the gel material is a solid with a finite shear modulus. (In a sample of vesicles without adsorbing polymer, the copper beads settled to the bottom of the vial.) The net force on the copper beads due to gravity is on the order of μN, so that each bead applied an average pressure of roughly 10 Pa, putting a very rough lower limit on the gel's yield stress.

**Computer simulations of nanoparticle binding** We carried out Brownian dynamics computer simulations of spherical nanoparticles binding to adhesive membranes to explore this system in microscopic detail and establish the mechanisms underlying its behavior. (Details are given in the Methods section and in SI). To parameterize the binding energy, we defined $w$ as the adhesion free energy per unit area. Because the deformation arises from a competition of adhesion *vs.* membrane-bending energies (as suggested in prior theory (6, 34)), we express adhesion as a dimensionless ratio, $wa^2/\kappa$, where $a$ is the nanoparticle radius and $\kappa$ is the membrane bending modulus (8) ($8.2 \times 10^{-20}$ J, appropriate for DOPC (35)).

Figure 7 shows the steady-state configurations obtained with increasing particle-membrane adhesion. When $wa^2/\kappa < 0.5$, simulations showed that particles adhered to the membrane without membrane tubulation or destruction. In the regime where $wa^2/\kappa > 0.8$, the simulations show a trend that ranged from partial buds to tubules to membrane-rupture as the particle area fraction was increased. The tubules were formed when a cluster of two or more particles was enveloped by the membrane. (See Supplemental Fig. S9 for snapshots of typical trajectories leading to tubule formation.) In the intermediate regime, $0.5 < wa^2/\kappa < 0.7$, simulations showed that the ruptured state was pre-empted by linear arrays of particles. It may be that these linear-array states represent a steady state or they might eventually nucleate tubules as suggested previously (25).

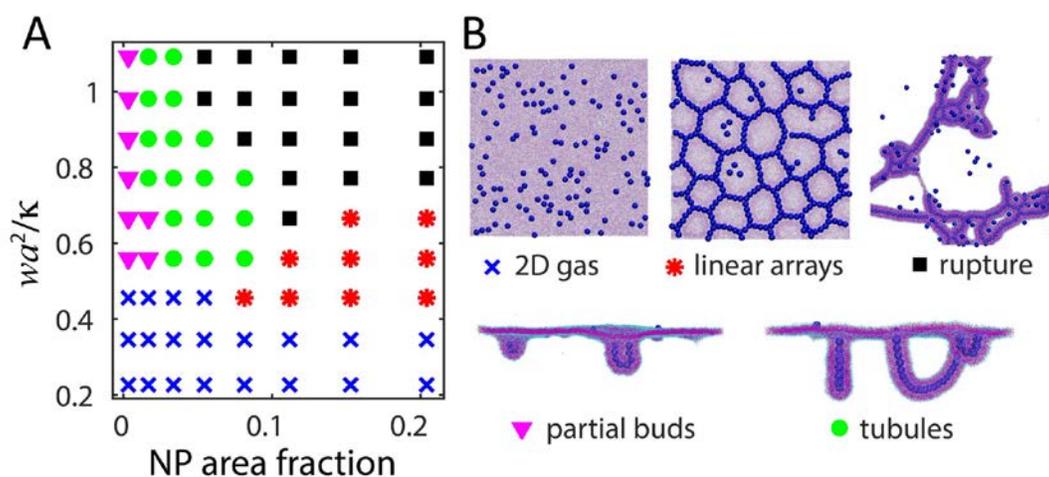

**Fig. 7**. (*A*) Diagram showing the steady-state configurations found in simulations as functions of dimensionless adhesion free energy and particle concentration. The symbols correspond to the states illustrated in (*B*). Particles are rendered in dark blue and membrane headgroups are in violet.

In our experiments, the particle area fraction was not fixed, but most likely increased over time as more particles bound to vesicles. The simulations' trend of partial buds, tubules and rupture with increasing particle density therefore correspond closely with the observed process of invagination (tubule formation) and pore formation over time in the experiments.

A key result of the simulations is a well-defined value of $wa^2/\kappa$ that defines a crossover from binding to tubule formation and rupture. In the following section we discuss the crossover from weak or partial wrapping to full wrapping of the particles as the trigger for the crossover.

**Discussion: Proposed mechanisms of vesicle adhesion, destruction, and the crossover** Vesicle adhesion and destruction were triggered by the adsorption of the nanoparticles, not by osmotic stress (see SI for control experiments). The role of the DOPS fraction in the membrane was to tune the adhesion strength between particles and the membrane by means of an electrostatic double-layer attraction between the cationic trimethyl ammonium on the Au-TTMA nanoparticles and the anionic phosphate on the DOPS lipid. Even in the absence of DOPS, we still observed nanoparticle binding, consistent with earlier findings that DOPC vesicles are anionic, with a zeta potential of –9 mV (electrophoretic mobility with 0.1 mM NaCl; (36)) and that they adhere to cationic particles (21, 37). Below, we quantitatively estimate $w$, the interaction energy per area between particles and vesicles. Here, we point out that the observed threshold increased from approx. 4 mol% to 5 mol% as the salt concentration was raised from no added salt to 20 mM NaCl. This observation is qualitatively consistent with the Poisson-Boltzmann theory of charge-based interactions in suspension, which predicts that adding salt weakens the interaction, so that more DOPS should be needed to drive the crossover.

In continuum theory, a crossover between the weakly-bound and fully-enwrapped configurations for a single particle was predicted from the Helfrich model of the membrane, accounting for large-amplitude deformations where linear superposition fails (6, 38). When there is no tension in the membrane and the interaction is of short range, the crossover to envelopment is discontinuous and occurs when $wa^2/\kappa = 2$.

Tension in the membrane is expected to shift the discontinuous transition to higher $w$ (6), while a finite range of interaction softens the crossover (8). This threshold can be reduced below 2 if there are many nanoparticles, as suggested in earlier studies with three or more particles (27) as well as in studies of non-spherical particles (39, 40). Our computer simulations showed a crossover to tubulation and strong destruction at a considerably lower threshold, $wa^2/\kappa$ near 0.5 (Fig. 7). The linear particle arrays shown in Fig 7*B* and in ref. (22) suggest a simple approximation, in which a linear particle aggregate is treated as a long cylinder lying in the plane of the membrane. The energy of bending around a cylinder of radius $a$ is 4× smaller than for bending around a sphere of radius $a$ because the membrane curves only in one direction. In the continuum limit and with a finite concentration of bound nanoparticles, this continuum model predicts a wrapping threshold when $wa^2/\kappa = ½$. Although the continuum approximation may not be strictly valid since $a$ is comparable to the membrane width, this estimate is nonetheless quantitatively consistent with the simulations.

For the experimental DOPC/DOPS/Au-TTMA system, we estimated the value of $w$ using Poisson-Boltzmann theory for the electrostatic double-layer interaction free energy per area between a flat plane (the membrane) and a spherical particle. The Au-TTMA surface potential was 18 mV, obtained from measured electrophoretic mobility (32), and the membrane's potential was taken as the sum of the pure-DOPC potential (-9 mV (36)) plus the potential coming from a charge of $-e$ in each added DOPS molecule. For a particle radius $a = 3.4$ nm, $wa^2/\kappa = ½$ corresponds to a composition of 9 mol% DOPS. Treating the nanoparticle-membrane interaction in terms of adhesion per area is a crude approximation because the range of interaction (set by the Debye length of 3 nm or more) is comparable to particle size. Nonetheless, the reasonable agreement of the theory with the measured threshold DOPS composition gives us confidence of predictive control of this process, and further indicates that the electrostatic binding and discontinuous wrapping are the principal mechanisms.

When the DOPS content was low (< 4 mol% with no added NaCl), the nanoparticles bound to the membrane and were able to spread laterally. By contrast, when the DOPS content exceeded the threshold value, we propose that the nanoparticles were completely enveloped by the membrane (either individually or in clusters), such that no part of a nanoparticle was exposed at the outer surface of the bilayer. In this way, the membrane continually enveloped particles and left unbound membrane exposed. Such a process of continuous envelopment would continually generate in-plane strain and force an overall remodeling of the membrane shape.

As particles bind to the membrane, the projected surface area of the membrane shrinks because of the envelopment of each bound nanoparticle (illustrated in Fig. 8). This model explains why the area reduction rate depends on the local nanoparticle concentration (Fig. 3). Assuming that each nanoparticle-wrapping event reduces the projected membrane surface area by an amount equal to the surface area of the nanoparticle, $4\pi a^2$, a steady area-reduction rate of 40 $\mu m^2$/s on a 5,000-$\mu m^2$ membrane corresponds to a flux of roughly 50 particles/($\mu m^2$•s) binding to the membrane. If the flux of particles were limited by their diffusion through water, the flux would be given by $3\phi D/(4\pi a^3 R)$, where $D$ is the nanoparticle diffusion constant, $R$ is the vesicle radius, and $\phi$ is the volume fraction of nanoparticles. From the known concentration of added nanoparticles, we verified that the diffusion-limited flux is high enough to account for the measured rate of vesicle shrinkage.

We propose that as the effective surface area shrinks, the interior vesicle volume can only decrease at a rate limited by water permeation through the membrane. If the binding is too fast, then tension should build up in the membrane and eventually reach the lysis tension, at which point the membrane should form a pore. If the area shrinkage is slow, however, the water permeation can keep pace with the area reduction and the tension stays below lysis; in such cases no pore is expected. In our experiments, long-lasting pores were only observed in vesicles whose projected area shrank at a rate faster than 300 $\mu m^2$/s (Fig. 3*C*). Using the reported permeability of DOPC membranes (41), we estimated a crossover shrinkage rate of order 0.1 $\mu m^2$/s. This value is far below the measured value. This difference suggests that the membranes may be far more permeable to water than expected owing to particle binding. This conclusion is consistent with previous work (21).

Without nanoparticles, tension-induced pores close very rapidly (42), but with nanoparticles the pores are stabilized by the "pearl necklace" arrangement of nanoparticle-lipid clusters at the pore's rim (Fig. 5). Since we never observed more than one pore on any vesicle, we conclude that once a pore forms, it allows rapid exchange of fluid so that the membrane tension remains below the lysis threshold.

Above the threshold binding energy, spots formed (Fig. 4) owing to clustering of nanoparticles, most likely because of attractive forces induced by the membrane deformation. Simulations indicate that the membrane-mediated attraction between particles occurs when the particles are strongly bound and highly wrapped (26, 43). Recent experiments with micron-scale particles confirmed this effect: weakly bound, partially wrapped particles had negligible lateral interactions, while fully-wrapped particles attracted one another over a distance of 3 particle diameters owing to the membrane deformations (14). With many particles present, membrane-mediated attraction can lead to compact clusters or linear aggregates (22, 23) and tubulation (8, 24, 43), consistent with our simulations (Fig. 7).

To form inward-facing (invaginated) tubules (Fig. S7), the particles must reside on the concave surface of the tubule. It may be that this configuration reduces the bending energy needed to enwrap the particles. Our finding that particles remain on the concave side of the tubules in the simulations is consistent with this explanation (Fig. 7). Previous experimental (18-20) and numerical (8, 23, 25, 26) studies of spherical particles or viruses binding to vesicle also show a tendency toward tubules with the particles on the inner, concave surface. Alternatively, it is possible that particle binding leads to a contraction of the outer leaflet of the bilayer, which would then favor negative curvature. Previous studies of cationic and anionic particle binding to phosphatidylcholine (PC) lipid membranes, however, indicated that cationic particles should tend to dilate the lipid layer (44, 45), which would more plausibly lead particles to favor positive curvature (the convex surface). Whatever the mechanism, the tubules invaginated such that particles remained on the concave surface while still remaining on the exterior surface of the vesicle.

By contrast, earlier reports showed that nanoparticles (21) or proteins (46) that bind on the exterior leaflet *without* wrapping can drive tubules that extend *outward* from the vesicle. This latter case was explained by a lateral pressure arising from steric interactions among the particles or proteins, leading to a dilation of the outer leaflet, which then forms the convex (outer) surface of the tubule. The previous experimental system (21) consisted of cationic particles with DOPC lipids (*i.e.*, in the weak-binding regime) but with high enough particle concentration to induce the lateral pressure. Hence, the change from outward-growing tubules (21) to our inward-growing tubules can be explained as consequence of the membrane envelopment that occurs with large $wa^2/\kappa$.

Why would the vesicles turn inside out? As tubules grow into the vesicle interior, they raise the interior pressure. This pressure is apparent in the time-series showing forcible ejection of an encapsulated lipid-based particle (Fig. S6). When a pore opens, the interior pressure forces the tubules to emerge as shown in Fig. 2 and illustrated in Fig. 8. In the final configuration, the particles still reside on the concave surface of the membrane tubules. In this state, however, the leaflet of the membrane that was initially on the interior (luminal) side ends up on the exterior side.

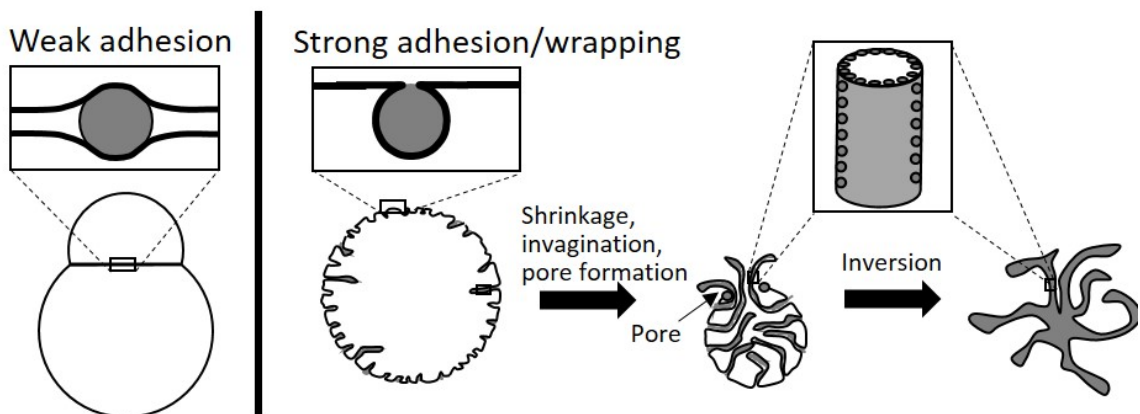

Fig. 8. Illustration of the adhesion and destruction/inversion processes.

We anticipate a similar inversion in any scenario where small spherical particles are added to the exterior of vesicles, provided that the binding energy exceeds the threshold value and the membrane fully wraps the particles. On the other hand, if such particles were added to the interior of vesicles, the same logic would predict outward-growing tubules and possibly a pore, but no inversion (consistent with earlier reports (18, 19)).

**Conclusions**

In our experiments, we exposed lipid bilayer membranes to cationic nanoparticles to understand how nanoparticle adhesion can be used to reshape the bilayer surface, a mechanism that could potentially be used to design novel responsive materials. We have successfully developed a membrane-particle system with highly tunable interactions, leading to the ability to form an adhesive network of vesicles (a bulk gel) or to drive a remarkable, catastrophic inversion of each vesicle leading to a network of tubules.

The crossover between the adhesion/gel regime and the destruction regime was driven by the tendency for particle adhesion to deform the membrane. When charged polymers (rather than spherical particles) were added, there was no membrane deformation and the gel structure was observed. With spherical nanoparticles, the crossover from adhesion to destruction/inversion occured at approximately 4% mole fraction DOPS. This result is consistent with the finding from simulations that destruction occurs when $wa^2/\kappa > ½$. Using Poisson-Boltzmann theory to estimate the double-layer interaction between the nanoparticles and membrane, we predict a threshold at 9 mol% DOPS. Given the level of approximation in treating the interactions, we regard this as a satisfactory agreement.

The gel that we found at low $w$ is a macroscopically large aggregate of vesicles that form a cohesive, closed-cell network. These gels are more than 99% water. Their morphology is reminiscent of cellular tissue, but is unusual among synthetic systems. Since the individual vesicles remain intact within the gel, they should be able to encapsulate multiple species in solution inside the gel. We envisage forming two different sets of vesicles, each one encapsulating a different reagent; the vesicles could then be dialyzed, mixed, and then made to form a vesicle gel. The two different species of reagent would not react with one another until the gel is ruptured in some way, causing their release.

When the DOPS concentration reached the threshold value of approx. 4 mol%, nanoparticles were fully enveloped by the membrane, causing the vesicle membrane to be loaded with adhered nanoparticles and ultimately causing destruction of the vesicle. The inversion/destruction regime results in complete and irreversible delivery of the contents of the vesicle. These results may lead to vesicles that are tailor-made to rupture and release only in response to selected particles (that bind strongly) and not to others. Such a system could be very useful for delivery in myriad contexts.

The results obtained with this tunable system suggest a unified model that could explain the wide variety of behaviors reported previously with vesicles plus nanoparticles and polymers. Under conditions of matched osmotic strength (as here), the deformations are caused by particle binding and membrane deformation. The key parameters are binding energy per area, $w$ (mediated by charge density on particles, on membranes, and in solution), which separates adhesion from destruction. We found that nanoparticle concentration and membrane permeability also play an important role: if the particles bind fast enough then the membrane can form a long-lasting pore.

**Materials and Methods**
Giant unilamellar vesicles (GUVs) were prepared using the electroformation technique first described by Angelova *et al*., (47) and later adapted and studied in detail by Herold, *et al*. (48). The majority lipid used in these experiments was the mono-unsaturated 1,2-dioleoyl-*sn*-glycero-3-phosphocholine (DOPC; Avanti Polar Lipids). The head group is nominally zwitterionic but measurements by Needham and co-workers showed that DOPC vesicles are slightly negative, with a (zeta potential of –9 mV in 0.1 mM NaCl (36). To add a controlled additional amount of negative charge, we used 1,2-dioleoyl-*sn*-glycero-3-phospho-L-serine (DOPS; Avanti Polar Lipids), which has an anionic head group (charge -*e*). To adjust the adhesion energy between the cationic nanoparticles and the membrane surface, several sets of vesicles were prepared, ranging from 100 mol% DOPC to 85 mol% DOPC + 15 mol% DOPS. For imaging fluorescence in some cases, we added a small amount of headgroup-labeled lipid 1,2-dioleoyl-sn-glycero-3-phosphoethanolamine-N-(lissamine rhodamine B sulfonyl) (ammonium salt); (Rh-DOPE; Avanti Polar Lipids). All vesicles reported here were formed in a 175 mOsm/L sucrose solution and then diluted with an equal volume of 180 mOsm/L glucose solution to make the vesicles slightly floppy before the Au-TTMA nanoparticles were added. In some cases (and only where explicitly stated), a controlled amount of NaCl was also added to the exterior solution to test for electrostatic effects.

For large-scale production of vesicles, we used Phospholipon 85G lipid and a gentle hydration method. Phospholipon 85G was used as received from the American Lecithin Company. It contains 91.5% PC-headgroup lipid, 2.8% Lyso-PC lipid, 1% unspecified nonpolar lipids, 0.3% PE lipid, and trace amounts of other non-lipid species. Lipid dissolved in chloroform was dried onto a clean glass surface (interior of a glass tube or the surfaces of glass microscope slides), dried in a vacuum chamber, then exposed to the sugar solution and placed in an oven at 35-40° for 24-48 h. By this method, total vesicle volumes as large as 50 mL were obtained in a single batch. These vesicles were then diluted with a 180 mOsm/L glucose solution, and allowed to sit for up to 1 day for the vesicles to settle. To make the gel, we added a solution containing 0.1% wt/vol poly-L-lysine (150 kDa) or PDADMAC (200 kDa), 5 mL of 175 mOsm/L sucrose, and 5 mL of 215 mOsm/L glucose. We verified that the solutions were osmotically matched with the vesicles.

The nanoparticles used in these experiments have a gold core and are functionalized with surface ligands consisting of a thioalkyl tetra(ethylene glycol)ated trimethylammonium (TTMA) ligand (Fig. 1*A*) (33). The tetra(ethylene glycol) spacer was added to keep the particles stable in suspension. Particles were synthesized using the Brust-Schiffrin two-phase synthesis method (49) and then functionalized with TTMA ligands *via* place exchange reactions (50). The core diameter was 2 nm (transmission electron microscopy), the hydrodynamic diameter was $6.7 \pm 0.4$ nm (dynamic light scattering), and the zeta potential in suspension was $18.2 \pm 0.8$ mV (electrophoretic mobility) (32).

The process of mixing vesicles and nanoparticles was monitored *in situ* using optical microscopy so that the early stages of adsorption could be visualized. To this end, we first added vesicles into a long, narrow perfusion chamber (Grace Bio Labs), then placed the chamber on the microscope. (See SI for more information.) We waited a few minutes to allow the vesicles to settle onto the coverslip. Then, we added 5 μL of the stock nanoparticle solution (approx. 1 mM of nanoparticles plus approx. 175 mOsm/L of glucose + sucrose with osmolarity checked) into one end of a perfusion chamber. As the nanoparticles diffused from one end of the sample chamber to the other, a visible 'front' of adhesion events was tracked across the sample. Differential interference contrast, bright-field, or dark-field images were acquired using a CoolSnap HQ2 camera (Roper Scientific) and a Zeiss 63× Plan Neofluar objective with 1.4 NA. Confocal images were obtained on a Marianas spinning disk confocal system (3I, Inc., Denver, CO), consisting of a Zeiss Observer Z1 microscope equipped with a Yokagawa CSU-X1 spinning disk confocal head, a QuantEM 512SC EMCCD camera, Plan Apochromat 63× or 100× oil immersion objective (1.4 NA) and Slidebook software.

We performed molecular dynamics simulations to determine how the particle-membrane adhesion strength changed dynamics and the steady-state configuration. We represented the membrane by the coarse-grained solvent-free membrane model developed by Cooke and Deserno (51), which is computationally tractable while capturing the relevant features of biological membranes. The lipids were represented by a linear polymer formed by three beads, one bead for the head and two beads for the tails. There are short-ranged attractive interactions between pairs of tail beads that represent hydrophobic effects, and short-range repulsions between pairs of head beads and head-tail pairs. The bending modulus, $\kappa$, was set to $8.2\times10^{-20}$ J, which is close to the value for DOPC.

We modeled nanoparticles as beads of radius $a = 5$ nm, roughly consistent with the gold nanoparticles used in the experiments. In our simulations, nanoparticles and membrane-head beads interacted through a Lennard-Jones potential, with well-depth $\varepsilon_{att}$ determining the strength of the nanoparticle-membrane attraction (which was tuned by salt concentration or lipid composition in the experiments). To represent excluded volume, there were also repulsive interactions between nanoparticles and lipid tail beads and nanoparticle-nanoparticle pairs.

Membranes were initially planar, approximating the fact that in the experiments the radii of curvature of the initial vesicles was much greater than $a$. We initialized a 170×170 nm membrane in the center of a box of height 150 nm. Tension was held near zero. We initialized $n$ nanoparticles in the upper half of the box, so that the nanoparticle area fraction (if all nanoparticles adsorbed) was given by $\rho_{\text{np}} = n\pi a^2/L^2$, where $L$ is the lateral membrane dimension. Periodic boundary conditions applied in the plane of the membrane, ensuring that nanoparticles remained on one side of the membrane (unless it ruptured).

**Acknowledgments**
We thank Tom Powers, Adrian Parsegian and Joel Cohen for helpful discussions. We acknowledge the NSF-funded Materials Research Science and Engineering Center (MRSEC) on Bioinspired Soft Materials (DMR-1420382). IKT acknowledges support through the Bio and Soft Matter Research Training (B-SMaRT) REU site at UMass (DMR- 1359191). VR acknowledges the NIH (EB022641). MFH and GRL acknowledge the NIH Award Number R01GM108021 from the National Institute of General Medical Sciences. Computational resources were provided by the NSF through XSEDE computing resources (XStream, Maverick, Bridges, Comet) through award number MCB090163 and the Brandeis HPCC, which is partially supported by the Brandeis MRSEC on Bioinspired Soft Materials (DMR-1420382).

**Cited References**
1.  van Meer, G., Voelker, D. R. & Feigenson, G. W. (2008) *Nat. Rev. Mol. Cell Biol.* **9,** 112.
2.  Zimmerberg, J. (2006) *Current Biol.* **16,** R272.
3.  Mellor, H. (2010) *Biochim. Biophys. Acta-Mol. Cell Res.* **1803,** 191-200.
4.  Richards, D. M. & Endres, R. G. (2017) *Rep. Prog. Phys.* **80,** 126601.
5.  Johannes, L., Parton, R. G., Bassereau, P. & Mayor, S. (2015) *Nat. Rev. Mol. Cell Biol.* **16,** 311-321.
6.  Deserno, M. (2004) *Phys. Rev. E* **69,** 031903.
7.  Le Bihan, O., Bonnafous, P., Marak, L., Bickel, T., Trepout, S., Mornet, S., De Haas, F., Talbot, H., Taveau, J. C. & Lambert, O. (2009) *J. Struct. Biol.* **168,** 419.
8.  Bahrami, A. H., Raatz, M., Agudo-Canalejo, J., Michel, R., Curtis, E. M., Hall, C. K., Gradzielski, M., Lipowsky, R. & Weikl, T. R. (2014) *Advances in Colloid and Interface Science* **208,** 214-224.
9.  Spangler, E. J., Upreti, S. & Laradji, M. (2016) *J. Chem. Phys.* **144,** 044901.
10. Ruiz-Herrero, T., Velasco, E. & Hagan, M. F. (2012) *J. Phys. Chem. B* **116,** 9595-9603.
11. Chen, X. M., Tian, F. L., Zhang, X. R. & Wang, W. C. (2013) *Soft Matter* **9,** 7592-7600.
12. Chen, L. P., Xiao, S. Y., Zhu, H., Wang, L. & Liang, H. J. (2016) *Soft Matter* **12,** 2632-2641.
13. Santhosh, P. B., Velikonja, A., Perutkova, S., Gongadze, E., Kulkarni, M., Genova, J., Elersic, K., Iglic, A., Kralj-Iglic, V. & Ulrih, N. P. (2014) *Chem. Phys. Lipids* **178,** 52-62.
14. van der Wel, C., Vahid, A., Saric, A., Idema, T., Heinrich, D. & Kraft, D. J. (2016) *Sci Rep* **6,** 32825.
15. Dietrich, C., Angelova, M. & Pouligny, B. (1997) *J. Phys. II* **7,** 1651-1682.
16. Koltover, I., Radler, J. O. & Safinya, C. R. (1999) *Phys. Rev. Lett.* **82,** 1991.
17. Ramos, L., Lubensky, T. C., Dan, N., Nelson, P. & Weitz, D. A. (1999) *Science* **286,** 2325.
18. Yu, Y. & Granick, S. (2009) *J. Am. Chem. Soc.* **131,** 14158.
19. Gozen, I., Billerit, C., Dommersnes, P., Jesorka, A. & Orwar, O. (2011) *Soft Matter* **7,** 9706-9713.
20. Ewers, H., Romer, W., Smith, A. E., Bacia, K., Dmitrieff, S., Chai, W. G., Mancini, R., Kartenbeck, J., Chambon, V., Berland, L., Oppenheim, A., Schwarzmann, G., Feizi, T., Schwille, P., Sens, P., Helenius, A. & Johannes, L. (2010) *Nat. Cell Biol.* **12,** 11-U36.
21. Li, S. & Malmstadt, N. (2013) *Soft Matter* **9,** 4969-4976.
22. Saric, A. & Cacciuto, A. (2012) *Phys. Rev. Lett.* **108,** 118101.

23. Saric, A. & Cacciuto, A. (2013) *Soft Matter* **9,** 6677-6695.
24. Reynwar, B. J., Illya, G., Harmandaris, V. A., Muller, M. M., Kremer, K. & Deserno, M. (2007) *Nature* **447,** 461.
25. Saric, A. & Cacciuto, A. (2012) *Phys. Rev. Lett.* **109,** 188101.
26. Bahrami, A. H., Lipowsky, R. & Weikl, T. R. (2012) *Phys. Rev. Lett.* **109**.
27. Raatz, M., Lipowsky, R. & Weikl, T. R. (2014) *Soft Matter* **10,** 3570-3577.
28. Raatz, M. & Weikl, T. R. (2017) *Advanced Materials Interfaces* **4**.
29. Xiong, K., Zhao, J. Y., Yang, D. W., Cheng, Q. W., Wang, J. L. & Ji, H. B. (2017) *Soft Matter* **13,** 4644-4652.
30. Chaudhuri, A., Battaglia, G. & Golestanian, R. (2011) *Phys. Biol.* **8,** 046002.
31. Paasonen, L., Laaksonen, T., Johans, C., Yliperttula, M., Kontturi, K. & Urth, A. (2007) *J. Control. Release* **122,** 86-93.
32. Jiang, Y., Huo, S., Mizuhara, T., Das, R., Lee, Y.-W., Hou, S., Moyano, D. F., Duncan, B., Liang, X.-J. & Rotello, V. M. (2015) *ACS Nano* **9,** 9986-9993.
33. Phillips, R. L., Miranda, O. R., Mortensen, D. E., Subramani, C., Rotello, V. M. & Bunz, U. H. F. (2009) *Soft Matter* **5,** 607–612.
34. Lipowsky, R. & Dobereiner, H. G. (1998) *Europhys. Lett.* **43,** 219-225.
35. Nagle, J. F., Jablin, M. S., Tristram-Nagle, S. & Akabori, K. (2015) *Chem. Phys. Lipids* **185,** 3-10.
36. Needham, D. & Zhelev, D. (2007) in *Perspectives in Supramolecular Chemistry: Giant Vesicles*, eds. Luisi, P. L. & Walde, P. (John Wiley & Sons, Hoboken, NJ).
37. Wang, L. & Malmstadt, N. (2017) *J. Phys. D-Appl. Phys.* **50,** 415402.
38. Deserno, M. & Gelbart, W. M. (2002) *J. Phys. Chem. B* **106,** 5543.
39. Dasgupta, S., Auth, T. & Gompper, G. (2013) *Soft Matter* **9,** 5473-5482.
40. Dasgupta, S., Auth, T. & Gompper, G. (2014) *Nano Lett* **14,** 687-93.
41. Olbrich, K., Rawicz, W., Needham, D. & Evans, E. (2000) *Biophys. J.* **79,** 321-327.
42. Sandre, O., Moreaux, L. & Brochard-Wyart, F. (1999) *Proc. Natl. Acad. Sci. U. S. A.* **96,** 10591-10596.
43. Reynwar, B. J. & Deserno, M. (2011) *Soft Matter* **7,** 8567.
44. Li, Y. & Gu, N. (2010) *J. Phys. Chem. B* **114,** 2749-2754.
45. Wang, B., Zhang, L. F., Bae, S. C. & Granick, S. (2008) *Proc. Natl. Acad. Sci. U. S. A.* **105,** 18171-18175.
46. Stachowiak, J. C., Hayden, C. C. & Sasaki, D. Y. (2010) *Proc. Natl. Acad. Sci. U. S. A.* **107,** 7781.
47. Angelova, M. I., Soleau, S., Meleard, P., Faucon, J. F. & Bothorel, P. (1992) in *Trends in Colloid and Interface Science VI*, Vol. 89, pp. 127-131.
48. Herold, C., Chwastek, G., Schwille, P. & Petrov, E. P. (2012) *Langmuir* **28,** 5518-5521.
49. Brust, M., Walker, M., Bethell, D., Schiffrin, D. J. & Whyman, R. (1994) *J. Chem. Soc., Chem. Commun.* **7,** 801–802.
50. Hostetler, M. J., Templeton, A. C. & Murray, R. W. (1999) *Langmuir* **15,** 3782-3789.
51. Cooke, I. R., Kremer, K. & Deserno, M. (2005) *Phys. Rev. E* **72,** 011506.

# Nanoparticles Binding to Lipid Membranes: from Vesicle-Based Gels to Vesicle Inversion and Destruction

Derek A. Wood[a], Sarah Zuraw-Weston[a], Ian K. Torres[a], YiWei Lee[b], Li-Sheng Wang[b], Ziwen Jiang[b], Guillermo R. Lázaro[c], ShiYu Wang[d], Avital A. Rodal[d], Michael F. Hagan[c], Vincent M. Rotello[b], and Anthony D. Dinsmore[a]

[a]Department of Physics, University of Massachusetts Amherst; [b]Department of Chemistry, University of Massachusetts Amherst; [c]Department of Physics, Brandeis University; [d]Department of Biology, Brandeis University

## Supplementary Materials

In these supplemental materials, we provide additional detail. Specifically, we show a schematic of the sample cells used, we describe control experiments that test the role of osmotic stress on the vesicle behavior; and we provide images and plots of additional data that were referred to in the main text.

### 1. A schematic of the sample cells used:
A top-view schematic of sample chamber setup is shown in Fig. S1. We used CoverWell™ perfusion chambers purchased from (Grace Bio Labs) with #1½ cover glass, mounted on an optical microscope. Nanoparticles were added from the right side, after which they diffused further into the sample. This method allowed observation of the vesicles as the nanoparticles bound. Vesicles that were farther from the point of nanoparticle addition had a lower nanoparticle concentration.

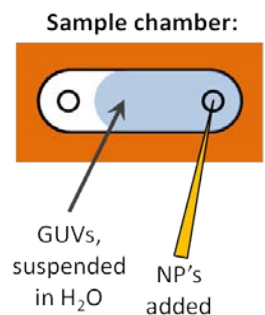

Fig. S1. Top view of perfusion chamber used for imaging the dynamics of nanoparticle/vesicle interactions.

### 2. Control experiments to test for osmotic stress effects:
We found that the critical molar ratio of 4.5% DOPS marking the crossover between the adhesion-only and vesicle-destruction regimes did not depend on the osmotic pressure imbalance between the vesicle interiors and the solution. Four samples of vesicles were prepared; vesicles with 4 mol% DOPS electroformed in 175 mOsm/L sucrose and diluted in 185 mOsm/L glucose with a 1:1 volume ratio (negative osmotic pressure, -10 mOsm/L), the same vesicles instead diluted with 165 mOsm/L glucose with a 1:1 volume ratio (positive osmotic pressure, 10 mOsm/L), and finally two more samples identical to the previous two but prepared with 5 mol% DOPS instead. Each sample was exposed to the same concentration of nanoparticles for 1 h, and in all cases the results matched those reported above (*i.e.*, 4 mol% led to adhesion and 5 mol% led to destruction).

### 3. Images show a sharp crossover from adhesion and gel formation to total destruction.
Figure S2 shows a representative set of optical images of samples with varying lipid composition, showing the steady state morphology. With no added NaCl, samples with average DOPS concentration was 3 mol% or less maintained a gel structure in steady state. In samples with 4 mol% or more, GUVs rapidly underwent destruction. With 4 mol% DOPS, a minority of vesicles survived in the steady state, whereas with 8 mol% DOPS, a negligible number of GUVs survived in the steady state. We attribute these surviving vesicles to variations in lipid composition of individual GUVs, so that a few individual vesicles may have been below the threshold DOPS fraction of 4%. With 5 mM or 10 NaCl, we could discern no change in the threshold DOPS fraction. With 20 mM NaCl, however, we found that the threshold increased to 5 mol% DOPS (Fig. S2).

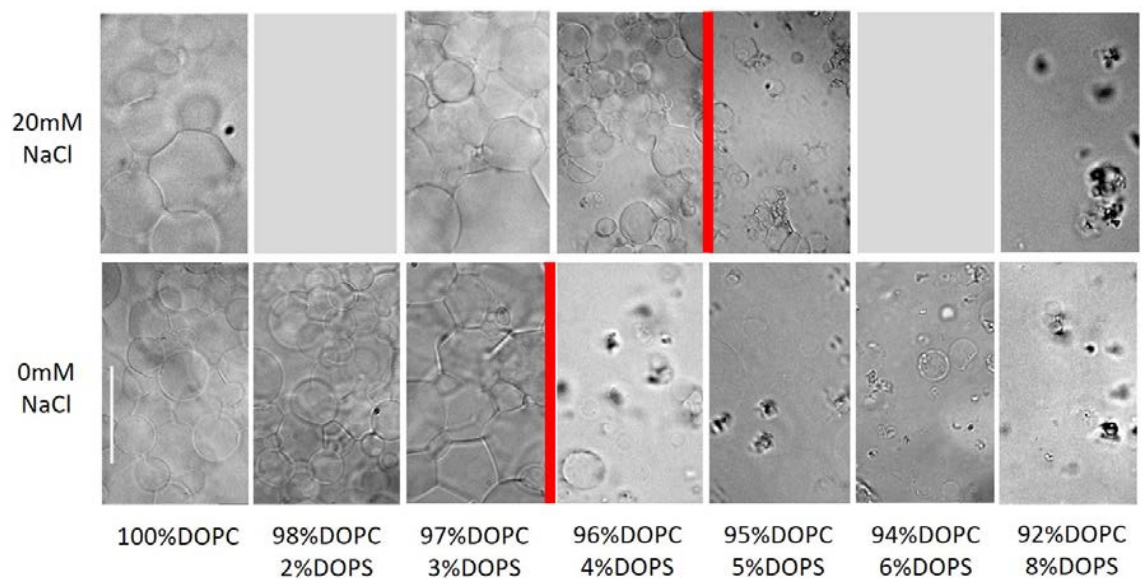

**Fig. S2**. Optical micrographs (bright field) of GUVs with varying lipid composition. The fraction of anionic DOPS increases from left to right. The scale bar is 20 μm and the scale is the same in all images. The red lines indicate the separation between samples that did or did not undergo vesicle destruction.

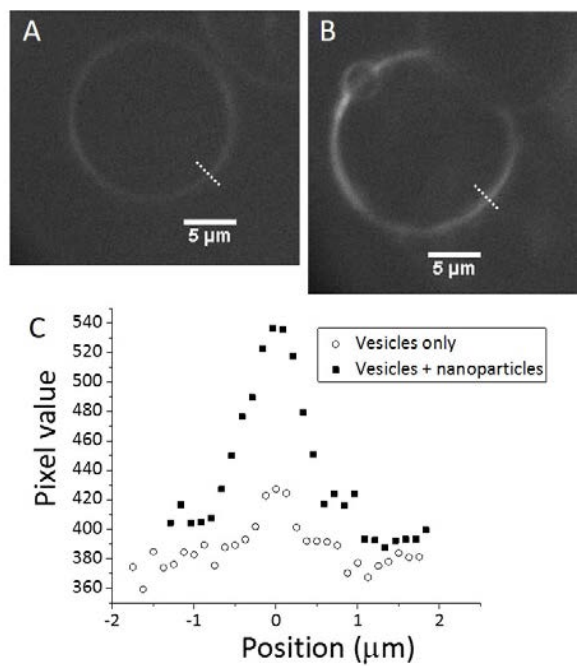

**Fig. S3.** (*A*) Dark-field image of a vesicle, showing faint contrast owing to light scattering from the membrane. (*B*) Dark-field image of a vesicle in the presence of nanoparticles, showing additional scattering by bound nanoparticles. (*C*) Plot of camera-pixel intensity *vs.* position along line segments shown by the white dashed lines in *A,B*. The scattering was enhanced by a factor of more than 3 by the bound nanoparticles.

## 4. Dark-field images showing nanoparticle binding:

Dark-field optical microscopy images indicate where the nanoparticles are concentrated. The image intensity comes from light that is scattered in the sample plane and the gold particles scatter much more strongly than lipids. Fig. S3 shows example dark-field images, providing evidence that the nanoparticles have bound at the vesicle's surface.

## 5. Bright-field images of multi-lamellar vesicles:

Many vesicles in each sample contain vesicles inside them. In such cases, we observed that the outermost membrane was 'attacked' by the nanoparticles, as shown in Fig. S4. In such cases, the outer lamellae were peeled back and remove from the vesicle one at a time.

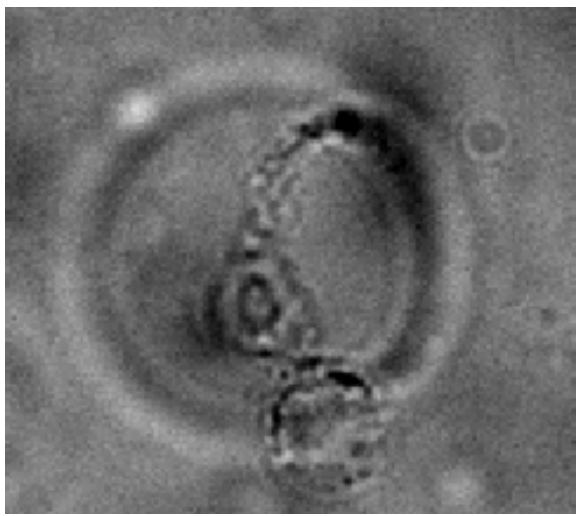

**Fig. S4.** Bright-field image of a multi-lamellar vesicle (5 mol% DOPS), in which the outermost lamella has been 'attacked' by nanoparticles and peeled away.

## 6. Images showing that the pore is open:

Figure S5 shows a sequence of bright-field microscope images of a vesicle containing many vesicles within it. In this case, the ejection of the interior vesicles proves that the pore is open. Figure S6 shows a sequence of images of a vesicle containing a lipid aggregate (or a multilamellar, onion-like vesicles) inside it. The interior aggregate diffused inside and was then ejected through the pore.

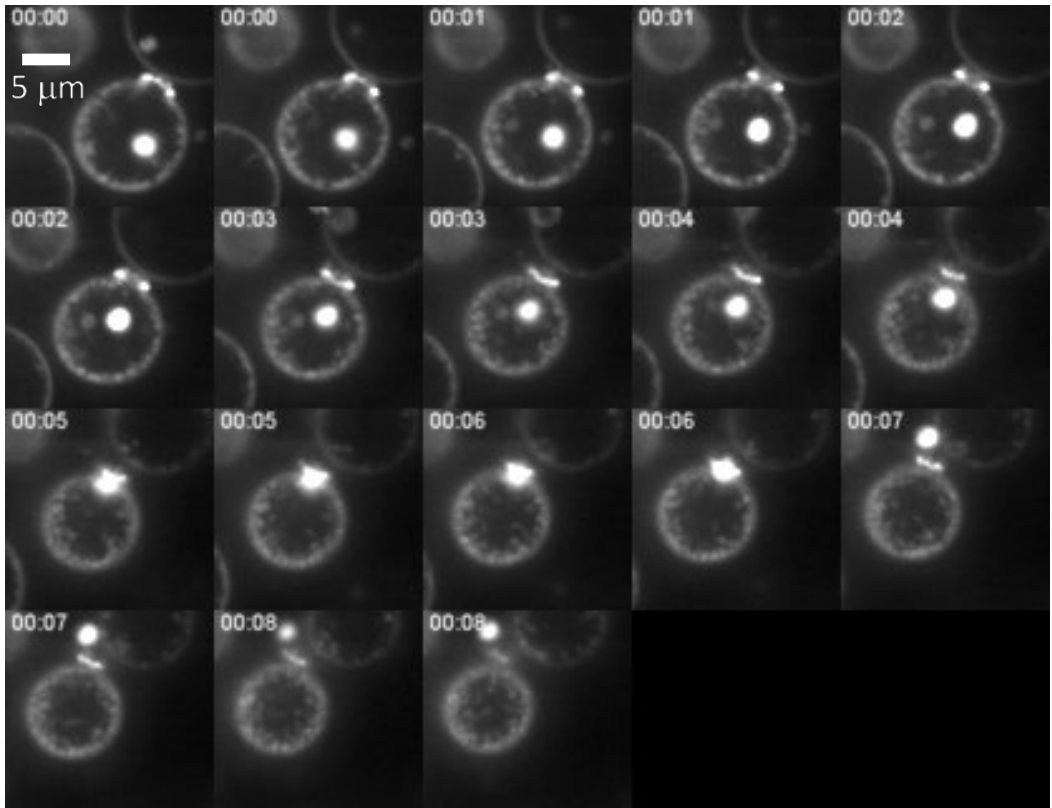

**Fig. S6.** A montage of images acquired with confocal fluorescence microscope. This vesicle contained 5 mol% DOPS + approximately 1 mol% Rh-DOPE and was exposed to Au-TTMA nanoparticles. Scale bar is provided in the first image. Initially, there was a large solid lipid-based object inside the vesicle. Over time, this object was forced out through the pore by the internal pressure. While this particle was inside the vesicle, it diffused slowly. It was then trapped in the pore for 3 frames, and then finally ejected a distance of more than 3 μm in the following frame ($t = 7$ s).

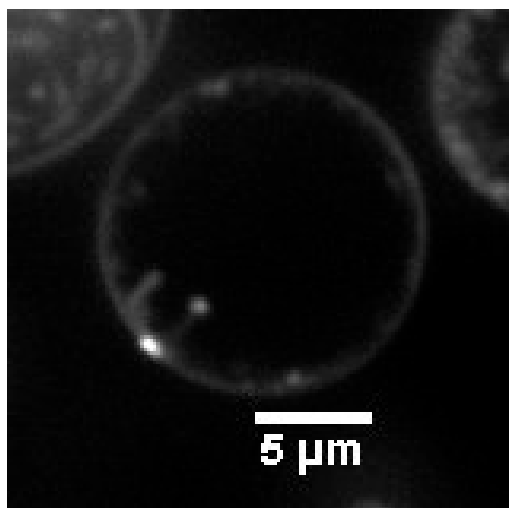

**Fig. S7.** A confocal microscope image of a vesicle containing 5 mol% DOPS + approximately 1 mol% Rh-DOPE, exposed to Au-TTMA nanoparticles. Inward-facing tubules (invaginations) are clearly visible in the image.

## 7. Confocal microscope images shows invaginated tubules.
Figure S7 shows a confocal fluorescence image of GUVs above the crossover DOPS fraction, in the destruction regime. The lipid is fluorescent and is seen to deform into the vesicle interior.

## 8. Bright-field images show the growth of an adhesion patch over time.
When the DOPS content was below the crossover value, we saw vesicles adhered to one another. Figure S8 shows a time-lapse sequence of images of the adhesion spots between vesicles.

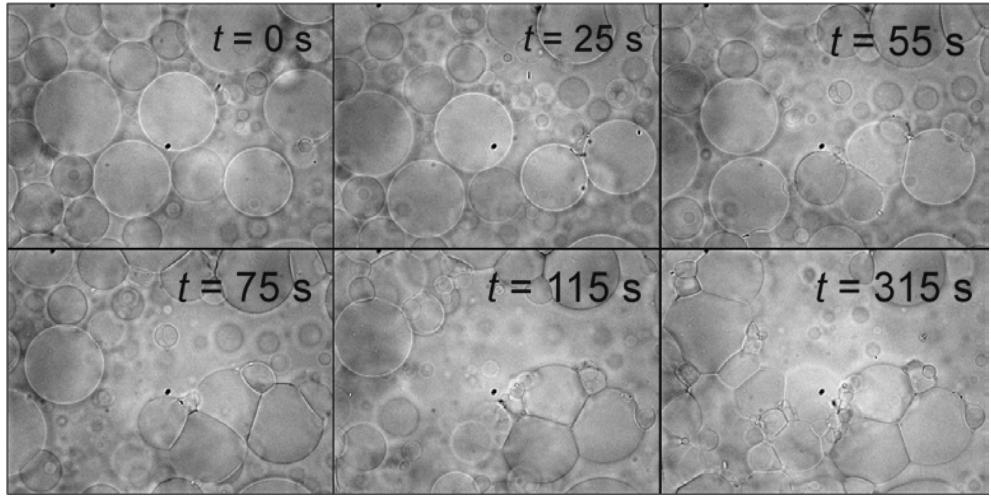

**Fig. S8.** Time lapse images show the adhesion process of DOPC vesicles (without DOPS) as Au-TTMA nanoparticles diffused into the imaged region from the right.

## 9. Simulation methods.
*Interaction potentials: Membrane model* – The three pseudoatoms in each model lipid were connected through two finite extensible nonlinear elastic (FENE) bonds,

$$U_{\text{bond}}(r) = -\frac{1}{2}k_{\text{bond}}r_{\text{cut}}^2\log[1-(r/r_{\text{cut}})^2].$$

with maximum bond length $r_{\text{cut}} = 1.5\sigma$ and force constant $k_{\text{bond}} = 30\epsilon_0/\sigma^2$, where the reference length $\sigma$ is the size of a lipid tail bead. Additionally, the first and third pseudoatoms were linked by a harmonic potential,

$$U_{\text{bend}}(r) = \frac{1}{2}k_{\text{bend}}(r-4\sigma)^2.$$

The excluded volume of membrane beads was represented by a Weeks-Chandler-Andersen (WCA) potential (1), with the interaction between two beads with indices $i$ and $j$ was given by

$$U_{\text{rep}}(r) = 4\epsilon_{\text{rep}}[(\frac{b_{i,j}}{r})^{12} - (\frac{b_{i,j}}{r})^6 + \frac{1}{4}],$$

with $\epsilon_{\text{rep}} = 1$ and cutoff $r_{\text{cut}} = 2^{1/6}b_{i,j}$. The parameter $b_{i,j}$ depends on the types of interacting beads $i$ and $j$: $b_{h,h} = b_{h,t} = 0.95\sigma$ and $b_{t,t} = 1.0\sigma$, with the subscripts `h' and `t' denoting head and tail beads, respectively. Hydrophobic interactions were captured by an attractive interaction between all pairs of tail beads:

$$U_{\text{hydro}}(r) = \begin{cases} -\epsilon_0, & r < r_c \\ -\epsilon_0\cos[\pi(r-r_c)/2\omega_c], & r_c \leq r \leq r_c + \omega_c \\ 0, & r > r_c + \omega_c \end{cases}$$

with $\epsilon_0 = 1.0$ and $r_c = 2^{1/6}\sigma$. The potential width $\omega_c$ determined, among other membrane properties, the membrane bending rigidity. We set it to $\omega_c = 1.7$ so that our bending modulus was $\kappa \approx 20k_BT$ (2).

*Interaction potentials: Nanoparticles* – We modeled nanoparticles as single beads of radius $a = 2.5$ nm. Nanoparticles interact with lipid tail beads and other nanoparticles through the repulsive component of the Lennard-Jones potential; *i.e.*, the interaction between beads $i$ and $j$ is

$$U_{\text{np,rep}}(r) = 4\epsilon_{\text{rep}}(\frac{b_{i,j}}{r})^{12},$$

with $\epsilon_{\text{rep}} = k_B T$ and cutoff radius $r_{\text{off}} = b_{i,j}$. For the interaction between nanoparticles and lipid tail beads, $b_{\text{np-tail}} = 3.0$ and for the interaction between pairs of nanoparticles $b_{\text{np-np}} = 5.0$. Nanoparticles experienced an attractive interaction with lipid head beads, represented by a Lennard-Jones potential; *i.e.*, the interaction between nanoparticle $i$ and head bead $j$ was

$$U_{\text{np}}(r) = 4\epsilon_{\text{att}}^*[(\frac{b_{i,j}}{r})^{12} - (\frac{b_{i,j}}{r})^{12}],$$

which was cut off at $r_{\text{off}} = 6.0$ and the parameter $\epsilon_{\text{att}}$ controlled the nanoparticle-membrane attraction strength. We estimate the adhesion free energy density from the interaction potential between lipid head beads and nanoparticles following Ruiz-Herrero, *et al.*(3),

$$\epsilon_a = -\gamma \log[1 + \sigma^{-1} \int_\sigma^\infty dr(e^{-U_{\text{np}}(r)} - 1)]$$

with $\gamma = 0.86/\sigma^2$ being the areal density of lipids.

*Simulations* – We performed simulations using HOOMD (4, 5). We considered a tensionless membrane by simulating in the NPT ensemble, allowing box changes in the *xy* directions to maintain a constant membrane tension. Membrane bead positions were propagated in time using the Martina-Tobias-Klein thermostat-barostat at $k_B T = 1.1\varepsilon$, $P_{xy}=0.0$, and coupling constants $\tau_T=0.5$ and $\tau_P=0.4$ for the thermostat and barostat, respectively. Nanoparticle positions were propagated using Brownian Dynamics. We performed simulations with a timestep length $\Delta t = 0.0025$, for $1.6\times10^7$ timesteps.

Snapshot images of tubules that formed in simulations are shown in Fig. S9. We discerned two kinds of tubules: U and I, as defined in the figure.

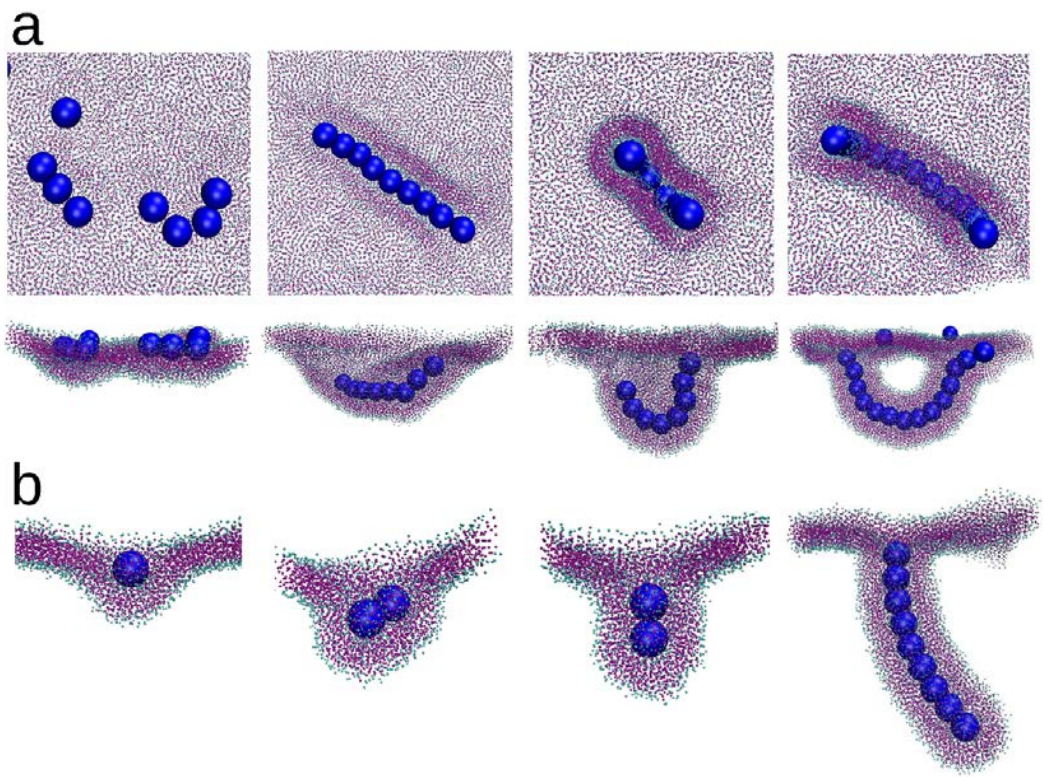

**Fig. S9**. Representative simulation trajectories illustrating the pathways of tubule formation. **(a)** Snapshots showing the formation of a 'U-tubule', meaning that the tubule is connected to the membrane at both ends. Particles initially formed a linear aggregate on the relatively flat membrane; subsequently the membrane wrapped the aggregate leading to tubulation. **(b)** Snapshots showing I-tubule formation, meaning that the tubule is connected to the membrane only at one end. Formation began with envelopment of two NPs, forming a duplet oriented normal to the membrane. The tubule then extended through diffusion and association of additional NPs.

## References for Supplemental Section

1. Weeks, J. D., Chandler, D. & Andersen, H. C. (1971) *J. Chem. Phys.* **54,** 5237.
2. Cooke, I. R., Kremer, K. & Deserno, M. (2005) *Physical Review E* **72**.
3. Ruiz-Herrero, T., Velasco, E. & Hagan, M. F. (2012) *J. Phys. Chem. B* **116,** 9595-9603.
4. Nguyen, T. D., Phillips, C. L., Anderson, J. A. & Glotzer, S. C. (2011) *Comput. Phys. Commun.* **182,** 2307-2313.
5. Anderson, J. A., Lorenz, C. D. & Travesset, A. (2008) *J. Comput. Phys.* **227,** 5342-5359.